\documentclass{iopconfser}

\usepackage{xcolor}

\begin{document}

\title{Classical and mixed classical-quantum systems from van Hove's unitary representation of contact transformations}

\author{Marcel Reginatto$^{1}$, Andr\'{e}s Dar\'{i}o Berm\'{u}dez Manjarres$^{2}$ and Sebastian Ulbricht$^{1,3}$}

\affil{$^1$Physikalisch-Technische Bundesanstalt, Bundesallee 100, 38116 Braunschweig, Germany}
\affil{$^2$Universidad Distrital Francisco Jos\'{e} de Caldas, Cra. 7 No. 40B-53, Bogot\'{a}, Colombia}
\affil{$^3$Institut für Mathematische Physik, Technische Universität Braunschweig,
 Mendelssohnstraße 3, 38106 Braunschweig, Germany}

\email{marcel.reginatto@ptb.de}

\begin{abstract}
Descriptions of classical mechanics in Hilbert space go back to the work of Koopman and von Neumann in the 1930s. Decades later, van Hove derived a unitary representation of the group of contact transformations which recently has been used to develop a novel formulation of classical mechanics in Hilbert space. 
This formulation differs from the Koopman-von Neumann theory in many ways.
Classical observables are represented by van Hove operators, which satisfy a commutation algebra isomorphic to the Poisson algebra of functions in phase space. Moreover, these operators are both observables and generators of transformations, which makes it unnecessary to introduce unobservable auxiliary operators as in the Koopman-von Neumann theory. 
In addition, for consistency with classical mechanics, a constraint must be imposed that fixes the phase of the wavefunction.
The approach can be extended to hybrid mixed classical-quantum systems in Hilbert space. The formalism is applied to the measurement of a quantum two-level system (qubit) by a classical apparatus.
\end{abstract}

\section{Introduction}

The most prominent formulation of classical mechanics in Hilbert space is the KvN theory, which is based on the insight of Koopman and von Neumann. They pointed out that the Liouville equation for the classical density in phase space, being a linear equation, may be re-expressed as an operator equation in an appropriately defined Hilbert space \cite{K1931,vN1931}. The first systematic effort to extend their work was done by Sudarshan \cite{S1976}, who was motivated by the measurement problem. He attempted to construct a Hilbert space theory of quantum systems interacting with classical measuring devices in which the classical sector of the theory was based on the KvN approach. However, his mixed classical-quantum theory encounters various difficulties \cite{PT2001,T2006}, in part due to some of the features of the KvN theory. While functions of phase space coordinates play a dual role in classical mechanics, that of observables and of generators of canonical transformations, in the KvN theory a phase space function is necessarily mapped to a \textit{pair} of different operators, depending on whether the operator should play the role of an observable or a generator. For example, in the KvN theory, the operator of time displacements is not an observable, while the energy observable does not generate the time evolution. As a result, the algebraic structure of classical mechanics that is contained in the Poisson algebra of classical observables does not have a unique counterpart in KvN theory. A further difficulty is the need to suppress superpositions as they are absent for classical systems, which is accomplished by super-selection rules, turning the theory into a hidden variable theory \cite{S1976}. Additional issues related to the proper handling of the phase of the classical wavefunction also arise \cite{S1976,M2002,GM2004,BGT2019}.

The alternative approach to classical mechanics in Hilbert space that we present in this paper takes as its starting point van Hove's unitary representation of the group of contact transformations, summarized in Section \ref{sec:vHr}. In van Hove's groundbreaking thesis \cite{vH1950, vH1951}, he showed that one may associate a unique operator with a function of phase space coordinates, such that the commutator algebra of the operators is isomorphic to the Poisson algebra of the functions in phase space. However, the definition of appropriate physical states requires some care: we have found that the phase of the classical wave function has to be subject to certain consistency requirements \cite{BRU2024}, described in Section \ref{Sec:CMvH}, that allows us to \textit{fix} the functional form of the phase of the classical wave function (see however Refs. \cite{BGT2019,GT2020} for a different approach which does not rely on fixing the phase). This approach that we develop in this paper leads to an alternative formulation of classical mechanics in Hilbert space that overcomes the difficulties of the KvN theory.
In addition, the algebraic structure of the operators allows for clear criteria to distinguish whether a theory is classical or quantum in nature, as we elucidate in Section \ref{sec:classicality}.
Furthermore, in Section \ref{Sec:H}, we discuss how our formalism can be extended to a novel hybrid theory of mixed classical-quantum systems that satisfies stringent consistency conditions.

\section{The van Hove representation of the Poisson Lie algebra of classical observables} \label{sec:vHr}

The problem of finding a unitary representation $\mathcal{R}$ of the group of contact transformations was solved by van Hove \cite{vH1950,vH1951}, who showed that $\mathcal{R}$ is reducible and that it is the continuous sum of a family of irreducible representations $\mathcal{R}^{(\alpha)}$ that depend on a real parameter $\alpha$. 
The representation $\mathcal{R}^{(0)}$ is closely related to the KvN approach to classical mechanics \cite{vH1951} and differs strongly in its structure from the representations with $\alpha \ne 0$ that we consider in this paper.

\subsection{The van Hove operators and their connection to pre-quantization.}

We consider the operators corresponding to the irreducible representation $\mathcal{R}^{(\alpha)}$ for $\alpha=1/\hbar$. Then, given a function $F(\mathbf{q},\mathbf{p})$ in phase space and its associated Hamiltonian vector field $\xi_F =\{F,\,\cdot\,\}$, we introduce the linear function $\theta(F)=F-\mathbf{p}\cdot\nabla_p F$ and define the van Hove operator \cite{vH1951}
\begin{equation}\label{vH_op}
\hat{\mathcal{O}}_{F} = \theta(F) + i \hbar \xi_F =  F- \mathbf{p}\cdot\nabla_p F + i \hbar \left(\nabla_q F\cdot \nabla_p-\nabla_p F\cdot \nabla_q\right).
\end{equation}
One can show that the commutator algebra of the van Hove operators is isomorphic to the Poisson algebra of the corresponding functions in phase space \cite{vH1951,BRU2024},
\begin{equation}\label{Iso_vH_CM}  [\hat{\mathcal{O}}_{F},\hat{\mathcal{O}}_{G}]=i\hbar\hat{\mathcal{O}}_{\{F,G\}}.
\end{equation}
In particular, this implies $[\hat{\mathcal{O}}_{q_j},\hat{\mathcal{O}}_{p_k}]=i\hbar\hat{\mathcal{O}}_{\{q_j,p_k\}}=i\hbar\delta_{jk}$, thus the operators corresponding to the position and momentum do not commute, unlike in KvN theory.  

As demonstrated by Groenewold and van Hove, there is no isomorphism between the commutator algebra of quantum operators and the algebra of Poisson brackets. Therefore, an equation of the form of Eq. (\ref{Iso_vH_CM}), while valid for van Hove operators, cannot hold for quantum mechanical operators \cite{vH1951,G1999}.  While often seen in a negative light, this no-go theorem is very useful as it gives us the tools to distinguish classical and quantum systems in a fundamental, algebraic way.

If the Hamiltonian vector field $\xi_F$ is complete, which is a requirement that van Hove imposed in his work \cite{vH1951}, then $\hat{\mathcal{O}}_{F}$ is essentially self-adjoint. However, following Gotay \cite{G1999,G2000}, we will not restrict our considerations to operators associated with classical observables whose Hamiltonian vector fields are complete. It will be useful to relax this requirement, but this does not mean that we may assume that a symmetric operator $\hat{\mathcal{O}}_{F}$ which is not essentially self-adjoint will always be acceptable as a physical observable. We will consider an example of this situation in Section \ref{Sec:timeOperator} when we discuss a classical time operator. 

The results of van Hove, further elaborated by Segal, Souriau, and Konstant, provided the basis for what later became known as pre-quantization \cite{G2000}. In geometric quantization, further steps are then carried out to go from the pre-quantization to a full quantum theory. In that context, the operators (\ref{vH_op}) are know as pre-quantum operators. In this paper, the focus will be on taking the insight of van Hove that leads to pre-quantization to construct a consistent theory of classical mechanics based on these operators.

\subsection{Ensembles on phase space as an alternative formulation of classical mechanics based on the operators of van Hove}

In addition to the formulation in Hilbert space described in the previous section, one may define a theory of \textit{ensembles on phase space} \cite{BRU2024} based on the van Hove operators of Eq. (\ref{vH_op}). For any function in phase space $F(\mathbf{q},\mathbf{p})$, we define observables in terms of the functionals
\begin{equation}\label{obsEPS}
    \mathcal{O}_{F}[\varrho,\sigma]:=\langle \phi|\hat{\mathcal{O}}_{F}|\phi\rangle=\int d\mathbf{q} d\mathbf{p}\,\bar{\phi}\,\hat{\mathcal{O}}_{F}\,\phi
\end{equation}
and, introducing the notation $\{\cdot,\cdot\}_{\varrho,\sigma}$ to distinguish functional Poisson brackets from the usual Poisson brackets of functions in phase space, we obtain an algebra of observables in terms of the functional Poisson brackets 
   $\{\mathcal{O}_{F},\mathcal{O}_{G}\}_{\varrho,\sigma}:=\int d\mathbf{q} d\mathbf{p} \left(\frac{\delta \mathcal{O}_{F}}{\delta \varrho}\frac{\delta \mathcal{O}_{G}}{\delta \sigma}-\frac{\delta \mathcal{O}_{F}}{\delta \sigma}\frac{\delta \mathcal{O}_{G}}{\delta \varrho}\right)$.
It can be shown that the algebra of the observables given by Eq. (\ref{obsEPS}) is isomorphic to the Poisson algebra of the corresponding functions in phase space \cite{BRU2024},
\begin{equation}
  \{\mathcal{O}_{F},\mathcal{O}_{G}\}_{\varrho,\sigma}=\mathcal{O}_{\{F,G\}},  \label{eqn:algebra}
\end{equation}
and, therefore, is also isomorphic to the commutator algebra of van Hove operators (a similar result holds for ensembles on configuration space \cite{H2008,HR2016}). Thus, everything that can be done in the Hilbert space formulation can be reformulated using ensembles on phase space \cite{BRU2024}. This has some advantages, such as, e.g., the ability to introduce constraints that go beyond the usual Dirac-type constraint in Hilbert space, which are restricted to the form $\hat{\mathcal{C}}|\phi\rangle = 0$ for some operator $\hat{\mathcal{C}}$.

\section{Classical mechanics of non-relativistic particles in terms of van Hove operators}
\label{Sec:CMvH}

In what follows, we use van Hove's ideas, as exposed in the last section, to give a Hilbert space formulation of classical mechanics that captures all the essential algebraic aspects of Hamiltonian mechanics thanks to the one-to-one correspondence between phase space functions $f(\mathbf{q},\mathbf{p})$ and van Hove operators $\hat{\mathcal{O}}_{f(\mathbf{q},\mathbf{p})}$.

With the algebraic structure of Eq. (\ref{eqn:algebra}) at hand, the next step is to construct the physical states, which are now represented by classical wavefunctions $\phi(\mathbf{q},\mathbf{p})$. It is necessary to ensure that the interpretation given to the amplitude and phase of the wavefunction is consistent with a classical interpretation. As we show below, this means that we need to impose conditions on the \textit{phase} of the wavefunction.

\subsection{Hamiltonian operator and the solution of the equations of motion}
\label{SubSec:H_operator_eqs_motion}
While the ideas presented in the previous sections are quite general and can be applied to various problems of classical physics, in what follows we restrict our discussion to a non-relativistic system of classical particles with the Hamiltonian $H=\frac{1}{2m}\mathbf{p}^2+V(\mathbf{q})$ and the Lagrangian $L=\frac{1}{2m}\mathbf{p}^2-V(\mathbf{q})$. The corresponding van Hove Hamiltonian operator is given by
\begin{equation}\label{vH_op_H}
\hat{\mathcal{O}}_{H} = \left( V(\mathbf{q}) -  \frac{1}{2m}\mathbf{p}^2\right) +  i\hbar\left( \nabla_q V \cdot \nabla_p - \frac{\mathbf{p}}{m} \cdot \nabla_q  \right).
\end{equation}
Writing the wavefunction in terms of Madelung variables, $\phi=\sqrt{\varrho}\,e^{i\sigma/\hbar}$, and evaluating the real and imaginary parts of $\bar{\phi}\, i\hbar \frac{\partial \phi}{\partial t}=\bar{\phi}\,\hat{{\mathcal O}}_H\phi$, we obtain decoupled equations for $\varrho$ and $\sigma$ \cite{BRU2024,GT2022}, 
\begin{equation}\label{rovHsvH}
\frac{\partial\mathcal{\varrho}}{\partial t}=\nabla_{q}V\cdot\nabla_{p}\varrho  -\frac{\mathbf{p}}{M}\cdot\nabla_{q}\varrho,\label{rovH} \qquad
	\varrho \, \frac{\partial\sigma}{\partial t} = \varrho \left[ \frac{\mathbf{p}^{2}}{2M}-V+\nabla_{q}V\cdot\nabla_{p}\sigma-\frac{\mathbf{p}}{M}\cdot\nabla_{q}\sigma \right].\label{svH}
\end{equation}
To interpret $\varrho$ and $\sigma$, it is convenient to rewrite Eqs. (\ref{rovHsvH}) in terms of $H$ and $L$ and Poisson brackets,
\begin{eqnarray}
\frac{\partial\mathcal{\varrho}}{\partial t}+\{\varrho,H\} &=& \quad\frac{d\varrho}{dt} \quad=\quad 0, \label{LE}\\
\varrho \left(  \frac{\partial\sigma}{\partial t}+\{\sigma,H\} \right) &=& \varrho \, \frac{d\sigma}{d t} ~\quad=\quad \varrho L,\label{dsigmadt}
\end{eqnarray}
where we introduced the total derivative, which for a function $F(\mathbf{q},\mathbf{p})$ is defined by $\frac{\partial F}{\partial t}+\{F,H\} = \frac{dF}{dt}$. Eq.~(\ref{LE}) implies that $\varrho$ must be interpreted as a \textit{classical density} that satisfies the Liouville equation, while Eq. (\ref{dsigmadt}) implies that $\sigma$ must be interpreted as the \textit{classical action}. This interpretation gives rise to \textit{consistency conditions} which must be imposed on the phase of the wavefunction: the last equality of Eq. (\ref{dsigmadt}) requires that the differential for $\sigma$ satisfies $d\sigma = \mathbf{p} \cdot d\mathbf{q} - Hdt$, as it holds for the classical action, when evaluated over a classical trajectory. This, by comparing to the left side of Eq. (\ref{dsigmadt}), implies that 
\begin{equation}\label{constr}
    \nabla_q \sigma =\mathbf{p},\quad \nabla_p \sigma=0,\quad \frac{\partial \sigma}{\partial t}=-H, 
\end{equation}
must hold along any classical trajectory. The first two constraints must be satisfied at any given time $t$, while the third one involves explicitly the time $t$. 

As we already discussed in a previous publication \cite{BRU2024},  Eqs. (\ref{dsigmadt}) and (\ref{constr}) {\it together} specify how to \textit{fix} the functional form of the classical action, i.e., the phase of the classical wavefunction. It is given by
\begin{equation}\label{sigmaConstr}
		\sigma(\mathbf{q},\mathbf{p},t) =  \eta(\mathbf{q},\mathbf{p}) + H(\mathbf{q},\mathbf{p})[\tau(\mathbf{q},\mathbf{p})-\tau(\mathbf{q'},\mathbf{p'})-t],
	\end{equation}	
	where $\eta$ and $\tau$ satisfy $\{\eta,H\}=L$ and $\{\tau,H\}=1$, and $\tau(\mathbf{q'},\mathbf{p'})$ is chosen so that the numerical value of $\tau$ satisfies $\tau=\tau(\mathbf{q'},\mathbf{p'})+t$. One can check directly that this choice of $\sigma$ satisfies Eq. (\ref{dsigmadt}). The functions  $\tau(\mathbf{q},\mathbf{p})$ and $\eta(\mathbf{q},\mathbf{p})$ always exist for integrable systems\footnote{ One can derive $\tau(\mathbf{q},\mathbf{p})$ by expressing the time parameter $t$ as a function of $\mathbf{q}$ and $\mathbf{p}$, while $\eta(\mathbf{q},\mathbf{p})$ can be determined from  $d\eta=\mathbf{p} \cdot d\mathbf{q}-H\left(\nabla_q \tau \cdot d\mathbf{q} + \nabla_p \tau \cdot d\mathbf{p}\right)$, as $\nabla_q\eta=\mathbf{p} -H\nabla_q \tau$ and $\nabla_p\eta= -H\nabla_p \tau$ lead immediately to $\{\eta,H\}=L$. }. A proof that this choice of $\sigma$ satisfies Eq. (\ref{constr}) is given in Appendix \ref{App:proofConstrSigma}. 
    
The issue of how to handle the phase of the classical wavefunction has been widely discussed for other formulations of classical mechanics in Hilbert space \cite{S1976,M2002,GM2004,BGT2019}. Since these formalisms fundamentally differ from our approach, it comes as no surprise that the proposed solutions are quite different in nature.
In Section \ref{Sec:Conclusions}, we will elaborate on this in more detail.

Having determined the phase of the classical wave function, we can associate to it any density $\varrho$ that solves the Liouville equation (\ref{LE}). It is sometimes convenient to represent $\varrho$ for a given Hamiltonian $H$ as an integral over classical trajectories for $\mathbf{q}$ and $\mathbf{p}$ with initial conditions $\mathbf{q'}$ and $\mathbf{p'}$. Expressing them as functions $\mathbf{Q}(\mathbf{q'},\mathbf{p'},t)$ and $\mathbf{P}(\mathbf{q'},\mathbf{p'},t)$ which satisfy $\mathbf{Q}(\mathbf{q'},\mathbf{p'},t)=\mathbf{q}$, $\mathbf{Q}(\mathbf{q'},\mathbf{p'},0)=\mathbf{q'}$, $\mathbf{P}(\mathbf{q'},\mathbf{p'},t)=\mathbf{p}$  and $\mathbf{P}(\mathbf{q'},\mathbf{p'},0)=\mathbf{p'}$
any density $\varrho$ takes the form \begin{equation}\label{rhoTrajectories}
		\varrho(\mathbf{q},\mathbf{p},t)= \int d \mathbf{q'} d \mathbf{p'} \, \varrho(\mathbf{q'},\mathbf{p'},0)\delta(\mathbf{q'}-\mathbf{Q}(\mathbf{q'},\mathbf{p'},t))\delta(\mathbf{p'}-\mathbf{P}(\mathbf{q'},\mathbf{p'},t))
	\end{equation} 
    where $\varrho(\mathbf{q'},\mathbf{p'},0)$ is the density at time $t=0$.
    We will make use of this representation in our further discussion.

\subsection{Expectation values}
We now calculate the expectation value of the van Hove operator $\hat{\mathcal{O}}_{F}$, which needs to coincide with the average of the phase space function $F(\mathbf{q},\mathbf{p})$ with respect to the density $\varrho$. If we write $\phi$ it in terms of $\varrho$ and  $\sigma$, satisfying Eq. (\ref{constr}) when evaluated over classical trajectories, we get
\begin{equation}
    \langle \phi|\hat{\mathcal{O}}_{F}|\phi\rangle 
    =
    \left. \int d \mathbf{q} d \mathbf{p} \, \mathcal{\varrho}\left[F+\left(\nabla_q \sigma -\mathbf{p}\right) \cdot \nabla_p F  - \nabla_q F \cdot \nabla_p \sigma  \right]\right|_{\varrho(\nabla_q \sigma - \mathbf{p})=0, \varrho \nabla_p \sigma = 0} =  \int d \mathbf{q} d \mathbf{p} \,\mathcal{\varrho}F 
\end{equation}
which shows that $\langle \phi|\hat{\mathcal{O}}_{F}|\phi\rangle=\int d \mathbf{q} d \mathbf{p}\,\mathcal{\varrho}F$ as required. \textit{We emphasize that this equality is a consequence of the consistency conditions of Eq. (\ref{constr}).}
    
\subsection{The classical propagator}

The classical propagator $K$ allows us to evolve the wavefunction $\phi$. It can be expressed in terms of the classical trajectories $\mathbf{Q}(\mathbf{q'},\mathbf{p'},t)$ and $\mathbf{P}(\mathbf{q'},\mathbf{p'},t)$ defined in Section \ref{SubSec:H_operator_eqs_motion}, and $\sigma$ as given by Eq. (\ref{sigmaConstr}),
\begin{equation}		
		K(\mathbf{q},\mathbf{p},\mathbf{q'},\mathbf{p'},t) = \delta(\mathbf{q'}-\mathbf{Q}(\mathbf{q'},\mathbf{p'},t)) \delta(\mathbf{p'}-\mathbf{P}(\mathbf{q'},\mathbf{p'},t))
		\exp\{i[\sigma(\mathbf{q},\mathbf{p},t)-\sigma(\mathbf{q'},\mathbf{p'},0)]/\hbar\},
\end{equation}
as expected since the density $\varrho$ evolves along the vector field determined by the classical trajectories.
We can check that this expression is correct by evaluating the wavefunction at time $t$. Using Eq. (\ref{rhoTrajectories}), we get
		\begin{eqnarray}
		\phi(\mathbf{q},\mathbf{p},t) &=&\int d\mathbf{q}' d\mathbf{p}' \, K(\mathbf{q},\mathbf{p},\mathbf{q'},\mathbf{p'},t)\phi(\mathbf{q},\mathbf{p},0)\nonumber\\
		 &=& \int d\mathbf{q}' d\mathbf{p}' \,  \delta(\mathbf{q'}-\mathbf{Q}(\mathbf{q'},\mathbf{p'},t))  \delta(\mathbf{p'}-\mathbf{P}(\mathbf{q'},\mathbf{p'},t)) \exp\{i[\sigma(\mathbf{q},\mathbf{p},t)-\sigma(\mathbf{q'},\mathbf{p'},0)]/\hbar\}\nonumber\\
		 &~& \qquad \times~
		 \sqrt{\varrho(\mathbf{q'},\mathbf{p'},0)}\exp\{i \sigma(\mathbf{q'},\mathbf{p'},0) /\hbar\}\nonumber\\
		  &=& \sqrt{\varrho(\mathbf{q},\mathbf{p},t)}\exp\{i \sigma(\mathbf{q},\mathbf{p},t) /\hbar\},
	\end{eqnarray}
as required.
\subsection{Energy displacements and the classical time operator}
\label{Sec:timeOperator}

As is well known, for instance from Pauli's famous footnote in his ``General Principles of Quantum Mechanics''  \cite{P1933}, there is no self-adjoint time operator in quantum mechanics conjugate to the Hamiltonian operator. Its existence would imply a continuous energy spectrum and no ground state since a time operator generates arbitrary changes in the energy.  

In classical mechanics, however, we may introduce a phase space function $\tau(\mathbf{q},\mathbf{p})$ that satisfies $\{\tau,H\}=1$ so that its numerical value $\tau=\tau_0+t$ corresponds to the time used to parameterize the classical trajectories.  When the energy is bounded from below by a smooth minimum, some subtle issues arise if $\tau$ is used as a generator of canonical transformations. In particular, the curves induced in phase space are necessarily \textit{incomplete}. This incompleteness causes the corresponding van Hove operator $\hat{\mathcal{O}}_{\tau}$ to be non-Hermitian \cite{G1999,G2000}.

We illustrate this with the example of a one-dimensional harmonic oscillator, with $H=\frac{1}{2m}p^2+\frac{m \omega^2}{2}q^2$ and $\tau=\frac{1}{\omega}\tan^{-1}\left(\frac{m \omega q}{p}\right)$, which satisfy $\{\tau,H\}=1$. 
For any phase space function $F(q,p)$, the curves generated by $\tau$ are given by $\delta F = \{F,\tau\}\delta\lambda$, where $\lambda$ is the curve parameter.
For the oscillator, this leads to     
\begin{equation}
        q(\lambda) = q_0\sqrt{\frac{E_0-\lambda}{E_0}},\quad
        p(\lambda) = p_0\sqrt{\frac{E_0-\lambda}{E_0}},\quad
        H(\lambda) = E_0 - \lambda.
\end{equation}
for position, momentum, and the Hamiltonian.
We observe that the curves are incomplete: as $q$ and $p$ must remain real, the parameter $\lambda$ can not take any real value but is restricted to $-\infty < \lambda \le E_0$, also preventing the energy from becoming negative. 

Although demonstrated here for the harmonic oscillator, these results are valid for all smooth, bounded potentials that resemble a quadratic potential close to the minimum, since they only depend on the behavior of the curves close to the potential minimum.
We can now use van Hove's prescription and write down the operator 
    \begin{equation}
        \hat{\mathcal{O}}_{\tau} = \tau(q,p) + \frac{qp}{2H} + i\hbar\left(\frac{p}{2H}\frac{\partial}{\partial p}+\frac{q}{2H}\frac{\partial}{\partial q}\right),
    \end{equation}
that generates changes in the value of the energy of the oscillator and corresponds to the classical time operator. Notice that it becomes ill-defined as $H \rightarrow 0$, indicating that it can no longer be applied in this limit.

\subsection{Energy eigenstates}
 To define classical energy eigenstates, $\phi_E = \sqrt{\varrho_E} \, e^{i\sigma_E/\hbar}$, it is convenient to use the representation of $\varrho$ of Eq. (\ref{rhoTrajectories}).  To define a density that only has trajectories with energy $E$, we introduce the constraint $E=\frac{1}{2m}\mathbf{p'}^2+V(\mathbf{q'})$ in the integral and require that the initial density is of the form $\varrho_E(\mathbf{q'},\mathbf{p'},0)=\pi(\mathbf{q'},\mathbf{p'})\delta\left(\frac{1}{2m}\mathbf{p'}^2+V(\mathbf{q'})-E\right)$, where $\pi(\mathbf{q'},\mathbf{p'})$ is a non-negative phase space function, only restricted by the condition $\int d \mathbf{q'} d \mathbf{p'}\varrho_E(\mathbf{q'},\mathbf{p'},0)\stackrel{!}{=}1$. Then,
    \begin{equation}\label{phiEnergyEigenstate}
	   \varrho_E(\mathbf{q},\mathbf{p},t)= \int d \mathbf{q'} d \mathbf{p'} \,  \varrho_E(\mathbf{q'},\mathbf{p'},0)\delta(\mathbf{q}-\mathbf{Q}(\mathbf{q'},\mathbf{p'},t))\delta(\mathbf{p}-\mathbf{P}(\mathbf{q'},\mathbf{p'},t)).
    \end{equation}
Moreover, the phase of the wavefunction is given by the $\sigma$ of Eq. (\ref{sigmaConstr}), where we can now replace $H$ by $E$, since the density only includes trajectories with $H=E$, leading to
	\begin{equation}
		\sigma_E(\mathbf{q},\mathbf{p},t) =  \eta(\mathbf{q},\mathbf{p}) + E[\tau(\mathbf{q},\mathbf{p})-\tau(\mathbf{q'},\mathbf{p'})-t].
	\end{equation} 

\subsection{Eigenfunctions and eigenvalues for arbitrary operators}

Consider an observable, represented by a phase space function $F(\mathbf{q},\mathbf{p})$. If we prepare a system at time $t=0$ in a state where the observable has a given value, so that $F(\mathbf{q},\mathbf{p})=f$, the density at time $t=0$ will be of the form $\varrho_f(\mathbf{q'},\mathbf{p'},0)=\pi(\mathbf{q'},\mathbf{p'})\delta\left(F(\mathbf{q}',\mathbf{p}')-f\right)$ and it will evolve according to
\begin{equation}\label{phiFEigenstate}
		\varrho_f(\mathbf{q},\mathbf{p},t)= \int d \mathbf{q'} d \mathbf{p'} \,  \varrho_f(\mathbf{q'},\mathbf{p'},0)\delta(\mathbf{q}-\mathbf{Q}(\mathbf{q'},\mathbf{p'},t))\delta(\mathbf{p}-\mathbf{P}(\mathbf{q'},\mathbf{p'},t)).
	\end{equation}
The phase of the wavefunction is given again by the $\sigma$ of Eq. (\ref{sigmaConstr}).


\subsection{Absence of an uncertainty principle}

For simplicity, here we consider a one-dimensional system. One can check that
\begin{equation}
	[\hat{{\mathcal O}}_q,\hat{{\mathcal O}}_p] = \left(q+i\hbar\frac{\partial}{\partial p}\right)\left(-i\hbar\frac{\partial}{\partial q}\right)-\left(-i\hbar\frac{\partial}{\partial q}\right)\left(q+i\hbar\frac{\partial}{\partial p}\right) = i\hbar \label{vHxpc}
\end{equation}
as expected, because the commutator algebra of van Hove operators is isomorphic to the Poisson algebra of functions in phase space. However, as $[\hat{{\mathcal O}}_q,\hat{{\mathcal O}}_p]=i\hbar$, it would seem that Eq. (\ref{vHxpc}) would lead to an uncertainty principle that would be incompatible with the existence of classical solutions which are well localized in $q$ and $p$. This is not the case.

The absence of an uncertainty principle was derived in a previous publication \cite{BRU2024} and we refer the reader to this article for details. Here we would like to point out that the usual derivations of the quantum uncertainty principle fail in the classical case because of a \textit{crucial difference} between the quantum operators $\hat{Q}$, $\hat{P}$  and the van Hove operators $\hat{{\mathcal O}}_q$, $\hat{{\mathcal O}}_p$: while $\hat{Q} \hat{Q} = \hat{Q^2}$ and $\hat{P} \hat{P} = \hat{P^2}$, for the van Hove operators we have $\hat{{\mathcal O}}_q \hat{{\mathcal O}}_q \ne \hat{{\mathcal O}}_{q^2}$ and $\hat{{\mathcal O}}_p\hat{{\mathcal O}}_p \ne \hat{{\mathcal O}}_{p^2}$.	
A brief calculation leads to
\begin{equation}
	\hat{{\mathcal O}}_q \hat{{\mathcal O}}_q = q^2+2i\hbar q \frac{\partial}{\partial p} - \hbar^2 \frac{\partial^2}{\partial p^2}, \qquad
	\hat{{\mathcal O}}_p \hat{{\mathcal O}}_p =  -\hbar^2\frac{\partial^2}{\partial q^2},
\end{equation}
which shows that $\hat{{\mathcal O}}_q \hat{{\mathcal O}}_q$ and $\hat{{\mathcal O}}_p \hat{{\mathcal O}}_p$ are not van Hove operators as they involve second derivatives. Thus, they are not associated with \textit{any} classical observables. Yet, the derivation of the uncertainty relation requires interpreting $\langle\phi|\hat{{\mathcal O}}_q\hat{{\mathcal O}}_q |\phi\rangle$ and $\langle\phi|\hat{{\mathcal O}}_p\hat{{\mathcal O}}_p |\phi\rangle$ as expectation values of the square of the position and the square of the momentum, respectively, which is not true for the van Hove operators. Therefore, \textit{there is no uncertainty relation in the van Hove formulation of classical mechanics}.
The set of van Hove observables $\hat{{\mathcal O}}_F$ does not form a product algebra. Given $\hat{{\mathcal O}}_F$ and $\hat{{\mathcal O}}_G$, the only general way to get a third observable is through their commutator  $\frac{1}{i\hbar}[\hat{{\mathcal O}}_F,\hat{{\mathcal O}}_G]=\hat{{\mathcal O}}_{\{F,G\}}$.

\subsection{Absence of a superposition principle}
\label{SubSec:AbsenceUP}

If we were to define a classical wavefunction via a linear superposition, $\phi= \phi_1 + \phi_2$, the density and phase of $\phi$ would be given by
\begin{equation}
    \varrho = \varrho_1 + \varrho_2 + 2 \sqrt{\varrho_1 \varrho_2}\,\cos[(\sigma_1 - \sigma_2)/\hbar],\quad
    \sigma = \hbar  \tan^{-1} \left(\frac{\sqrt{\varrho_1} \sin(\sigma_1/\hbar)+\sqrt{\varrho_2} \sin(\sigma_2/\hbar)}{\sqrt{\varrho_1} \cos(\sigma_1/\hbar)+\sqrt{\varrho_2} \cos(\sigma_2/\hbar)} \right).\label{phiSuperposition}
\end{equation}
The expression for the phase in Eq. (\ref{phiSuperposition}) is obviously not of the form of Eq. (\ref{sigmaConstr}). Therefore, it will not satisfy the constraints on the phase of Eq. (\ref{constr}), which we must impose on all physically acceptable wavefunctions. Thus, we conclude that a superposition of two physically acceptable wavefunctions does not lead to another acceptable wave function. This means that \textit{there is no superposition principle in van Hove mechanics}.

Among other consequences, this implies the absence of interference effects, which are not observed for classical systems, as is very well known. Systems described by mixtures of states are not excluded, in agreement with classical mechanics.

Sudarshan achieves this result in KvN theory more indirectly \cite{S1976}, also concluding that it is impossible to have coherent superpositions of pure states, but that mixtures are allowed \cite{GM2004}. His approach relies on the definition of superselection operators and the associated superselection rules. In our formalism, it is not necessary to introduce such rules, since the absence of superpositions follows from requiring consistency via Eqs. (\ref{constr}).




\section{What is classicality? A precise algebraic definition } 
\label{sec:classicality}

One of the major difficulties of formulating hybrid systems is to find an appropriate definition of ``classicality'' that will allow us to unambiguously distinguish between the classical and quantum sectors of a given hybrid system, even if they are interacting. We will use an algebraic approach to deal with this issue.

To motivate our approach, it is useful to review Dirac's original proposal for quantization \cite{C2022}. It consists of requirements that turned out to be inconsistent, nevertheless, his proposal has led to fruitful discussions and useful developments. Dirac introduced four basic requirements, or {\it rules}, for any set of operators $\hat{\mathcal{D}}_F$ that are candidates for quantum operators representing a phase space function $F(\mathbf{q},\mathbf{p})$:
	\begin{enumerate}
		\item Linearity rule: $\hat{\mathcal{D}}_{a F + b G}=a \hat{\mathcal{D}}_{F} + b \hat{\mathcal{D}}_{G}$
		\item Power rule: $\hat{\mathcal{D}}_{F^n}=(\hat{\mathcal{D}}_F)^n$
		\item Identity rule: $\hat{\mathcal{D}}_1=\hat{1}$ 
		\item ``Poisson bracket $\rightarrow$ commutator'' rule: $[\hat{\mathcal{D}}_F,\hat{\mathcal{D}}_G] = i\hbar \hat{\mathcal{D}}_{\{F,G\}}$
	\end{enumerate}
Note that there is no universal agreement on the best way of formulating Dirac's rules, so they are sometimes formulated in a slightly different form or with additional requirements \cite{G1999,G2000,C2022}. Of relevance here is that neither the Schr\"{o}dinger operators $\hat{F}$, nor the van Hove operators $\hat{\mathcal{Q}}_F$ satisfy all four of these rules:
The Schr\"{o}dinger operators, on the one hand, satisfy rules 1 to 3 but not rule 4, an issue that is known as the Groenewold-van Hove theorem \cite{G1999,C2014}, which is a consequence of the commutator algebra of quantum operators being non-isomorphic to the Poisson algebra of functions in phase space \cite{G1999}. On the other hand,  the van Hove operators satisfy rules 1, 3, and 4 but not rule 2 because they do not form a product algebra, as we already discussed in Section \ref{SubSec:AbsenceUP}. Thus, while both sets of operators, classical and quantum, satisfy rules 1 and 3, they differ crucially because they satisfy different (Poisson or commutator) algebras. This provides a fundamental distinction between the observables of classical-quantum systems, their transformations via generators, and their dynamics, allowing us to distinguish classical systems from quantum ones. 

\section{Extension to hybrid systems }
\label{Sec:H}
We now extend the van Hove formulation of classical mechanics to accommodate \textit{hybrid systems} that consist of a classical system interacting with a quantum system. The basic idea is to describe the classical observables using van Hove operators and the quantum observables using Schr\"{o}dinger operators, both acting on a hybrid wavefunction in an appropriately defined Hilbert space.  Thus, classical observables will be given by van Hove operators $\hat{\mathcal{Q}}_F$ corresponding to functions $F(\mathbf{q},\mathbf{p})$ depending on the phase space coordinates $\mathbf{q}$ and $\mathbf{p}$  of a classical particle, while quantum observables $\hat{F}$ are Schrödinger operators depending on the configuration space coordinate $\mathbf{x}$ of the quantum particle. Interaction terms may involve all classical and quantum coordinates. 

\subsection{Definition of hybrid systems }
Let the wavefunction $\psi(\mathbf{q},\mathbf{p},\mathbf{x})=\sqrt{\varrho}\,e^{\,i \sigma / \hbar}$  describe the state of a classical particle of mass $m_C$ interacting with a quantum particle of mass $m_Q$ via a Galilean invariant potential $V(|\mathbf{q}-\mathbf{x}|)$. Following our recent publication \cite{BRU2024}, the equation of motion generated by the Hamiltonian operator of the hybrid system is given by 
\begin{equation}
    i\hbar\frac{\partial\psi}{\partial t} = \hat{{\mathcal O}}_{H} \psi = \left[-\frac{\hbar^{2}}{2m_Q}\nabla_x^{2}-\frac{\mathbf{p}^{2}}{2m_C} + V + i\hbar\left( \nabla_q V \cdot \nabla_p - \frac{\mathbf{p}}{m_C} \cdot \nabla_q  \right) \right]\psi.\label{psiHybrid}
\end{equation}
The corresponding equations for the density and phase are 
\begin{eqnarray}	\frac{\partial\mathcal{\varrho}}{\partial t}+\frac{\nabla_{x}\cdot\left(\mathcal{\varrho}\nabla_{x}\sigma\right)}{m_Q}+\nabla_{q}\varrho\cdot\frac{\mathbf{p}}{m_C}-\nabla_{p}\varrho\cdot\nabla_{q}V &=& 0,\label{rhoHybrid}\\	\frac{\partial\sigma}{\partial t}+\frac{\left|\nabla_{x}\sigma\right|^{2}}{2m_Q}-\frac{\hbar^{2}}{2m_Q}\frac{\nabla_x^{2}\sqrt{\mathcal{\varrho}}}{\sqrt{\mathcal{\varrho}}}+\nabla_{q}\sigma\cdot\frac{\mathbf{p}}{m}-\nabla_{p}\sigma\cdot\nabla_{q}V-\frac{\mathbf{p}^{2}}{2m}+V &=& 0.\label{sigmaHybrid} 
\end{eqnarray}	
Eqs.~(\ref{rhoHybrid})-(\ref{sigmaHybrid}) coincide with the Madelung form \cite{GT2020} of the hybrid system given in Ref. \cite{BGT2019}. However, while there is equivalence at the level of equations with the formalism presented in Ref. \cite{GT2020,BGT2019}, it is not clear to what extent both the interpretations of the formalisms and the actual calculations match. This will be discussed in a future publication. One possible reason for potential discrepancies is the difference in the definitions of physical quantities between the two approaches (such as the densities associated with the classical and quantum sectors and the definition of observables/generators). Another reason is that the phase of the hybrid wavefunction is treated differently. Eqs.~(\ref{psiHybrid})-(\ref{sigmaHybrid}) are treated according to our formalism in Ref. \cite{BRU2024} in more detail.
In Section \ref{Sec:example} of this work we give a brief example of a classical system interacting with a discrete quantum system, e.g., a two-level system or a qubit. 

\subsection{Locality and other consistency requirements }
We list various consistency conditions satisfied by our formalism, including locality (several of the proofs omitted here can be found in a previous publication \cite{BRU2024}). 
\begin{enumerate}
\item Conservation of probability. The probabilities of the classical and quantum sectors are given by marginalization, 
\begin{equation}	                 
\varrho_C(\mathbf{q},\mathbf{p})
=\int d \mathbf{x} \, \varrho(\mathbf{q},\mathbf{p},\mathbf{x}),\quad
\varrho_Q(\mathbf{x})=
\int d \omega \, \varrho(\mathbf{q},\mathbf{p},\mathbf{x}).
\end{equation}
\item Conservation of energy.
\item Galilean invariance for potentials of the form $V(|\mathbf{q}-\mathbf{x}|)$.
\item The two minimal conditions proposed by Salcedo \cite{CS1999,S2007} are satisfied. The commutators of the set of observables form a Lie algebra, and the algebra is isomorphic to the Poisson algebra for the van Hove operators and to the quantum commutator algebra for the Schr\"{o}dinger operators. 
\item The ``definite benchmark" that Peres and Terno \cite{PT2001}  proposed for ``an acceptable classical-quantum hybrid formalism'' is satisfied.
\item As there is no product algebra for the set of all observables, various no-go theorems in the literature \cite{CS1999,S1996,S2004} which assume such a product algebra and the Leibniz rule do not apply.
\item Strong separability. All classical observables are represented by van Hove operators $\hat{{\mathcal O}}_{F}$ which depend exclusively on coordinates $\mathbf{q}$ and $\mathbf{p}$, and they therefore commute with all quantum observables, which are represented by operators $\hat{G}$, which depend exclusively on coordinates $\mathbf{x}$. Thus, 
\begin{equation}
    [\hat{{\mathcal O}}_{F},\hat{G}]=0.
\end{equation}
As a consequence, a transformation acting solely on the quantum component cannot lead to changes in the expectation values of observables in the classical component and viceversa. This means that the hybrid theory does not allow for ``ghost interactions'' \cite{S2012}, nor does it permit non-local signaling.
\end{enumerate}

\subsection{An example: von Neumann measurement of a qubit with a classical apparatus }
\label{Sec:example}

As an application, we consider a measurement of a qubit by a classical apparatus. For simplicity, the classical apparatus is represented by a one-dimensional pointer. The model is based on the measurement scheme proposed by von Neumann \cite{vN1958}, which we generalize to our hybrid system. It involves a coupling between the momentum of the pointer and the quantum observable as well as the assumption that the interaction is only turned on for sufficiently short timescales. Therefore, we may ignore all the other contributions to the Hamiltonian. We allow for a classical potential $V_C(q)$ and consider an interaction term $\kappa p \hat{\sigma}_3$ with the standard Pauli matrix $ \hat{\sigma}_3$, where $\kappa$ is a coupling constant. Keeping in mind that $\hat{{\mathcal O}}_{p}=-i\hbar\frac{\partial}{\partial q}$, the hybrid equation of motion is given  by 
\begin{equation}\label{hybridWithQubit}
 i\hbar\frac{\partial\psi}{\partial t} = \hat{{\mathcal O}}_{H} \psi =\left\{ \left[-\frac{p^{2}}{2m_C} + V_C + i\hbar\left( \frac{\partial V_C}{\partial q} \frac{\partial}{\partial p} - \frac{p}{m_C} \frac{\partial}{\partial q}  \right)\right] + B_0 \hat{\sigma}_3 - \kappa i \hbar\frac{\partial}{\partial q} \hat{\sigma}_3   \right\}\psi . 
\end{equation}
Assuming a two-component wave function $\psi=(\psi_+,\psi_-)$ with $\psi_{\pm}=\sqrt{\varrho_{\pm}}\,e^{i \sigma_{\pm}/\hbar}$, Eq.~(\ref{hybridWithQubit})  is equivalent to the two equations
\begin{equation}
    i\hbar\frac{\partial\psi_{\pm}}{\partial t} =\left\{-\frac{p^{2}}{2m_C} + V_C + i\hbar\left( \frac{\partial V_C}{\partial q} \frac{\partial}{\partial p} - \frac{p}{m_C} \frac{\partial}{\partial q}  \right)  \pm B_0 \mp \kappa i \hbar\frac{\partial}{\partial q}  \right\}\psi_{\pm}\,.   
\end{equation}
We now follow the approximation of von Neumann and neglect all but the last term in the Hamiltonian while the interaction is turned on. This generates a displacement $q \rightarrow q \mp K$, where $K$ is the total displacement over the interaction time $T$.  

We assume that the position and momentum of the pointer are initially well localized so that the pointer's distribution is given by $\delta_{\epsilon}(q)\delta_{\epsilon}(p)$, where $\delta_{\epsilon}$ approximates a delta function. The probability of measuring one of the quantum states is $w_{\pm}$, with $w_{+}+w_{-}=1$, such that $\varrho_{\pm} = w_{\pm}\delta_{\epsilon}(q)\delta_{\epsilon}(p)$. It follows that the probability density of the pointer at time $T$ after the measurement is given by
\begin{equation}\label{pointer}
P(q,T) = \sum_{\pm} \int dp \,\varrho_{\pm}(q,p,T) = w_+\, \delta_{\epsilon}(q-K) + w_{-}\delta_{\epsilon}(q+K).
\end{equation}
We see that the initial probability density $P(q,0)=\delta_{\epsilon}(q)$ is displaced by $\pm K$ with probability $w_{\pm}$. The result is that the density of the pointer becomes correlated with the quantum observable. 

We now introduce the conditional density operator \cite{HR2016} and evaluate it after the measurement. We find
\begin{equation}
	\hat{\rho}_{Q|C}(T)
	=\int dq dp \,\left.\left(
	\begin{array}{cc}
		\varrho_{+} & \sqrt{\varrho_{+}\varrho_{-}}e^{i((\sigma_{+}-\sigma_{-})/\hbar)} \\
		\sqrt{\varrho_{+}\varrho_{-}}e^{-i((\sigma_{+}-\sigma_{-})/\hbar)} & \varrho_{-}
	\end{array}
	\right)\right|_{t=T}
	=\left(
	\begin{array}{cc}
		w_+ & 0 \\
		0 & w_-
	\end{array}
	\right),	
\end{equation}
since $\varrho_{+}(T)\varrho_{-}(T)=0$. Thus, if the pointer is initially localized and its position can be approximated by a delta function, the conditional density operator {\it decoheres} with respect to the $\hat\sigma_3$ basis after the measurement takes place. This result is in agreement with the one that is obtained using ensembles on configuration space \cite{HR2016}. 

\section{Summary and some concluding remarks}
\label{Sec:Conclusions}

Taking as our starting point the operators introduced by van Hove \cite{vH1951}, we formulate classical mechanics in Hilbert space. Functions of phase space coordinates, which play a dual role as both observables and generators of infinitesimal transformations in classical mechanics, are assigned to their corresponding van Hove operators. Defining physical states, however, requires some care as the phase of the classical wavefunction is subject to constraints to ensure consistency with its interpretation as the classical action, which follows from the equations of motion for the wavefunction expressed in terms of Madelung variables. These constraints, which are particular to classical mechanics and have no counterpart in quantum mechanics, are a crucial ingredient of our formalism. One important consequence of imposing the constraints is that the expectation values of the van Hove operators equal the average values of their associated phase space functions. Thus, in our approach, the van Hove operators play the dual role of observables and of generators, which is not the case in the KvN theory. We have explicitly worked out the example of a von Neumann measurement where a classical measuring apparatus interacts with a quantum two-level system (qubit). 

Since the commutator algebra of van Hove operators is isomorphic to the Poisson bracket algebra of phase space functions, the position and momentum operators are canonically conjugate and, hence, \textit{do not commute}. In the literature, it has been erroneously claimed that operators that represent classical position and momentum observables \textit{must} be commuting operators, as it is argued that non-commutativity implies an uncertainty relation. However, this assumption does not apply to our formalism. The reason is that the van Hove operators do not form a product algebra, which is needed to derive an uncertainty relation. This could have been expected since the operators already satisfy the other three rules that Dirac proposed for quantization (see Section \ref{sec:classicality}), and it is known that the full set of all four Dirac's rules is inconsistent.

The extension to mixed classical-quantum systems is straightforward and leads to a hybrid theory that satisfies consistency conditions such as the conservation of probability and energy, as well as various additional conditions proposed in the literature. For example, non-local signaling or so-called ``ghost interactions'' are not possible in the theory. Additionally, a commutator is defined for all observables, with the sets of classical and quantum observables being characterized by different Lie algebras.  

Our formulation of hybrid systems provides a counterexample to the assumptions of a large number of information-theory-based no-go theorems concerning hybrid systems, which all assume that the classical observables commute. Furthermore, any no-go theorems about hybrid systems that assume a product algebra do not apply to it.   


Finally, one can reformulate the Hilbert space theory of classical mechanics as a theory of ensembles on phase space, which is essentially a phase space version of the theory of classical ensembles on configuration space \cite{HR2016}. One can show that the three theories are equivalent as long as only classical systems are considered \cite{BRU2024}. 

\section*{Acknowledgments}
We thank Cesare Tronci and Michel Pannier for fruitful discussions, and the organizers of the DICE 2024 meeting where part of this work was presented. 
S.U. acknowledges funding by the Deutsche Forschungsgemeinschaft under Germany’s
Excellence Strategy EXC 2123 QuantumFrontiers Grant No.
390837967.

\appendix

\section{Proof that the $\sigma$ of Eq. (\ref{sigmaConstr}) satisfies Eq. (\ref{constr})}
\label{App:proofConstrSigma}

If we solve Eq. (\ref{dsigmadt}) with the $\sigma$ of Eq. (\ref{sigmaConstr}), we can derive the equality
\begin{equation}
\varrho \, \left[ \frac{\partial\sigma}{\partial t} + \{\sigma,H\}-L\right] = \varrho \left[\nabla_{q}\sigma\cdot\frac{\mathbf{p}}{m}-\nabla_{p}\sigma\cdot\nabla_{q}V-\frac{\mathbf{p}^{2}}{m} \right] = 0.   
\end{equation}
As it is valid for arbitrary choices of $\varrho$, it has to be true when we consider densities that consist of single trajectories described by delta functions, $\varrho(\mathbf{q},\mathbf{p},t;\mathbf{q'},\mathbf{p'})=\delta(\mathbf{q}-\mathbf{Q}(\mathbf{q'},\mathbf{p'},t))\delta(\mathbf{p}-\mathbf{P}(\mathbf{q'},\mathbf{p'},t))$. As the trajectories satisfy $\dot{\mathbf{q}}=\frac{\mathbf{p}}{m}$ and $\dot{\mathbf{p}}=-\nabla V$, we can write
\begin{equation}
\varrho \left[\left(\nabla_{q}\sigma-\mathbf{p}\right)\cdot\dot{\mathbf{q}}+\nabla_{p}\sigma\cdot \dot{\mathbf{p}} \right] = 0,  
\end{equation}
no matter which trajectory was chosen or the particular values of $\dot{\mathbf{q}}$ and $\dot{\mathbf{p}}$. It follows that $\varrho\left(\nabla_{q}\sigma-\mathbf{p}\right)=0$ and $\varrho\nabla_{p}\sigma=0$, which are the first two conditions of Eq. (\ref{constr}) (note that these are generally valid, as an arbitrary density that evolves in time is just a collection of single trajectories). Furthermore, if we evaluate the partial derivative of $\sigma$ of Eq. (\ref{sigmaConstr}) with respect to $t$, we immediately get $\partial \sigma / \partial t = -H$, and the third condition of Eq. (\ref{constr}) is also satisfied. 

Appendix A of Ref. \cite{BRU2024} provides an example showing how Eq. (\ref{constr}) is satisfied in the case of free fall.

\end{document}